# Influence of ns-laser wavelength in laser-induced breakdown spectroscopy for discrimination of painting techniques

Xueshi Bai[a,b], Delphine Syvilay[a,b], Nicolas Wilkie-Chancellier[b], Annick Texier[a], Loic Martinez[b], Stéphane Serfaty[b], Dominique Martos-Levif[c], Vincent Detalle[c*]

[a] LRMH, CRC USR 3224, 29, rue de Paris, 77420 Champs-sur-Marne, France

[b] SATIE, UMR CNRS 8029, Université de Cergy-Pontoise, site de Neuville rue d'Eragny, 95031 Neuville sur Oise, France

[c] C2RMF, UMR CNRS 171, 14, Quai François Mitterrand, 75001 Paris, France

**Abstract**

The influence of ns-laser wavelength to discriminate ancient painting techniques such as are fresco, casein, animal glue, egg yolk and oil was investigated in this work. This study was carried out with a single shot laser on samples covered by a layer made of a mixture of the cinnabar pigment and different binders. Three wavelengths based on Nd: YAG laser were investigated (1064, 532 and 266 nm). The plasma is controlled at the same electron temperature after an adjustment of pulse energy for these three wavelengths on a fresco sample without organic binder. This approach allows to eliminate the effects of laser pulse energy and the material laser absorption. Afterwards, the emission spectra were compared to separate different techniques. The organic binding media has been separated based on the relative emission intensity of the present CN or $C_2$ rovibrational emissions. In order to test the capability of separating or identifying, the chemometric approach (PCA) was applied to the different matrix. The different solutions in term of wavelength range to optimise the identification was investigated. We focused on the evaluation for the laser wavelength to insure a better separation. The different capacity was interpreted by differentiating the binders by the altered interaction mechanisms between the laser photon and the binders. Also, the electron temperature in the plasma was estimated, which provided the evidences to our findings.

---

[1] Corresponding author. E-mail address: vincent.detalle@culture.gouv.fr

# 1. Introduction

To appropriately conserve and restore the cultural heritage (CH) objects in the future, it requires to establish an adequate restoration strategy. Therefore, having detailed composition of the CH objects, ensure the products used to restore are easily removed for future treatment and will not cause further damage. Among the scientific analytical techniques used for material composition characterization, the laser technology based spectroscopic techniques are widely employed on cultural heritage for the research and development activities, including Raman spectroscopy [1-3], laser induced fluorescence (LIF spectroscopy) [4-8], laser ablation inductively coupled plasma mass spectrometry (LA-ICP-MS) [9-11and laser-induced breakdown spectroscopy (LIBS) [12-14]. From CH objects comprised of a vast variety of materials LIBS technique being straightforward, can be applied to all types of samples without any preparation. Moreover, LIBS provides an *in situ*, rapid, and micro-destructive approach for identification. [15, 16]

Previously our work identified wall paintings pigments with LIBS used multivariate treatments [17]. Other researchers have utilized LIBS to identify organic binders corresponding to different painting techniques. Overall, these studies present that LIBS can be used to analyse inorganic materials but also, the organic ones such as ancient restorative polymer products [18, 19]. The most commonly used light source of LIBS is the Q-switched Nd: YAG laser that operates at its fundamental (1064 nm) or harmonics wavelengths (532, 355 and 266 nm). The plasma is generated by absorption of laser energy in the material generally based on one of these interaction mechanisms: photothermal ablation, photochemical ablation and photophysical ablation, which depend on the photon energy [20]. Additionally, the absorption of different photons also relies on the diffuse reflectance of the pigments in the pictorial layer. [21] Also, the binding media in the layers can change the absorption of different excitation wavelengths. Therefore, studying if one laser wavelength can improve the ability of identifying the painting materials, the pigments and binders, etc. is important. In this study, we experimentally examined LIBS performances at different laser wavelengths for discriminating various ancient wall painting techniques. Then we compared the emission spectra from the plasma, particularly the molecular emission of CN and $C_2$ bands, which can be used as the mark the presence of organic material in the samples. [22] To advance the separation of unknown materials, chemometric method based on the principal component analysis (PCA) can be used. Moreover, analysing the property of the plasma can assist in understanding the mechanism during the generation of the plasma.

## 2. Experimental approach

### 2.1 Samples

The work has been achieved on the mock-up samples made of cellular concrete coated with lame plaster. The samples were then painted with a layer of the pigment cinnabar (HgS) mixed in the binder with different ancient painting techniques: fresco, casein, animal glue, oil and egg yolk (see Table 1). The samples used in this experiment are also one part of samples presented and characterized in our previous work [23]. The presence of proteins gives a high level of CN bonds, and long chain molecules contains C-C bond. They can also product carbon atoms by laser ablation which can be formed CN bands and $C_2$ bands with the chemical reaction in the plasma.[24, 25] The emission of such molecular bands can be used in LIBS signal analysis.

**Table 1 Binders and painting techniques used in this work and their description of composition. [26]**

| Binder/Technique | | Description |
|---|---|---|
| **Inorganic** | None /Fresco | The binder made by carbonation of calcium hydroxide containing lime ($CaCO_3$). |
| | Casein | Casein a milk protein isoforms (α, β and γ) are isolated from the precipitation of skim milk by an acid. |
| | Animal glue | Collagen is the main protein isolated from animal skin samples and contain high levels of amino acids such as glycine, proline and hydroxypoline. The paint layer is made by rearrangement of the collagen molecules. |
| **Organic** | Oil | Linseed oil (communly used) is a drying oil made up mainly of linoleic acid (15%) and linolenic acids (52%), which are polyunsaturated fatty acids. |
| | Egg yolk | Emulsion consisting of various molecules: water (51%), lipids (17-38%) and proteins (15%). The paint layer is formed by drying proteins (Dehydrating) |

## 2.2 Experimental setup and measurement conditions

The experimental setup is shown in Fig. 1. Two Nd:YAG lasers (Quantel) were used in the work. The first one (Ultra 100) provides on the fundamental of Nd:YAG at 1064 nm and the second one (CFR) generates the second harmonic at 532 nm and the forth harmonic at 266 nm. They all operated at a repetition rate of 20 Hz but selected one shot for one position insuring a good energy stability. The nominal pulse duration is 8 ns (FWHM). The pulse energies for each wavelength delivered on the sample surface was adjusted and controlled by attenuators and a power meter (not shown in the figure). The three laser beams were superposed on the same path in order to have a plasma at the same place for the detection system by two harmonic beam splitters: BS1 (Reflects 532 nm and transmits 1064 nm) and BS2 (Reflects 266 nm, transmits 532 nm and 1064 nm). A mechanical beam shutter (BSH) controlled the delivery of the single laser pulse onto the sample. Laser pulses were focused onto the sample with a lens (L1) of 30 cm focal length. Then we have put the focus plan under the sample surface in order to avoid the direct ionization in the air, so the diameter of craters for three laser wavelengths was about 200 μm.

During measurements, the sample was translated using X-Y stages in order to provide a fresh surface for each laser shot. The distance between the focusing lens (L1) and the sample surface was kept constant during the measurement by using a monitoring system which consisted in a combination of a laser pointer and three LEDs with their beams in oblique incidence onto the sample surface and the cross point on the focal plane (not shown in Fig. 1).

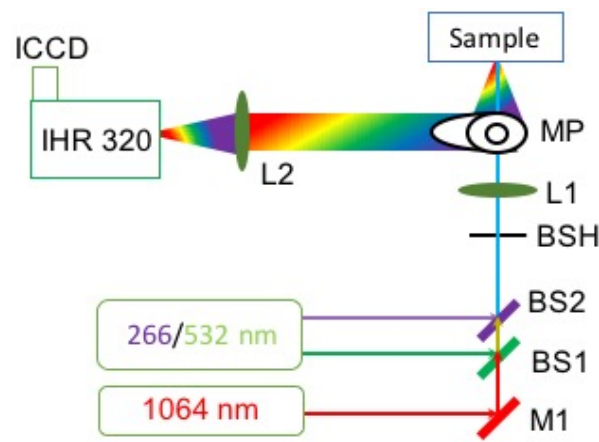

**Fig.1.** Schematic presentation of the experimental setup. M1: mirrors, L1, L2: lenses, BSH: beam shutter, BS1, BS2: beam splitter

The produced plasma emission was collected by off-axis parabolic mirror (parallel to focused beam) with a hole to permit the laser beam passing through and sending the plasma emission to the entrance of a spectrometer assisted with a lens (L2) with a focal length of 200 mm. A Czerny-Turner spectrometer (iHR320, HORIBA Scientific) with a grating of 300 lines/mm was coupled to the ICCD camera (Andor Technology, DH340T-18F-E3). By turning the grating with different center wavelength on the ICCD, the detection spectral range is from 250 nm to 812 nm (Table 2). With an entrance slit of 100 μm, the spectral resolution of the spectrometer was about 0.41 nm. Time-resolved detection was performed for spectroscopic detection by triggering the ICCD camera with the synchronization signal generated by a delay generator. The detection delay and the corresponding detection gate width used in the experiment were fixed for monitoring the plasma over a time interval from 100 to 2000 ns after the laser pulse (Parameters for ICCD: Delay: 100 ns; Gate: 1900 ns). In the experiment, each spectrum was acquired by one laser shot and the measurement was repeated 5 times over 5 ablation sites. The standard deviation was about 5% for the averaged intensity (maximum deviation on the whole spectrum) of these five spectra is measured.

**Table 2**

The detection spectral range for each grating position. (Unit: nm)

| Centre wavelength | 310 | 400 | 520 | 650 | 750 |
|---|---|---|---|---|---|
| Start wavelength | 245 | 330 | 453 | 580 | 680 |
| End wavelength | 370 | 465 | 586 | 712 | 812 |

To compare different capacities of different laser wavelength for discriminating different binders, the influence of different laser wavelengths must be eliminated in order to restrict the variable only on the binders. Therefore, first the laser pulse energy was adjusted to have the same plasma for each wavelength during the detection time interval. This allows the line emission intensities to be almost the same after adjusting the laser pulse energy. We chose the fresco mock-up sample because of its simple composition: pigment without organic binder, and took the calcium (Ca I 364.4 nm) from the matrix ($CaCO_3$), whose emission lines shown in Fig. 2. The adjusted excitation fluence of the laser pulse used in this study is shown in the Table 3. With these laser pulse energies, representing the total emission spectra for all grating detection with corresponding laser pulse energies are presented in Fig. 3. The element emission lines are identified for example mercury is a characteristic of cinnabar (435.8 nm) , calcium, carbon and potassium, etc.

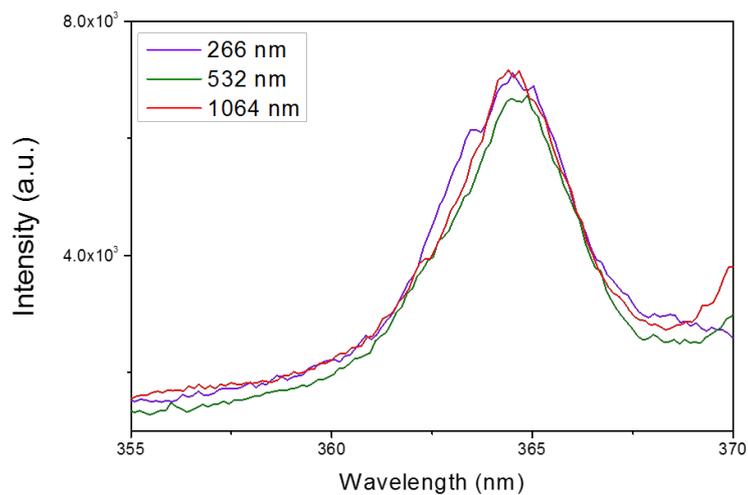

**Fig. 2.** The line emission of calcium (Ca I 364.4 nm) in the matrix (CaCO3) induced by three laser wavelengths after the laser energy adjustment

**Table 3**

The excitation fluence adjusted of the laser pulse used in this study

| Excitation wavelength (nm) | Excitation fluence (J/cm$^2$) |
| :---: | :---: |
| 266 | 4.5 |
| 532 | 5.5 |
| 1064 | 13.2 |

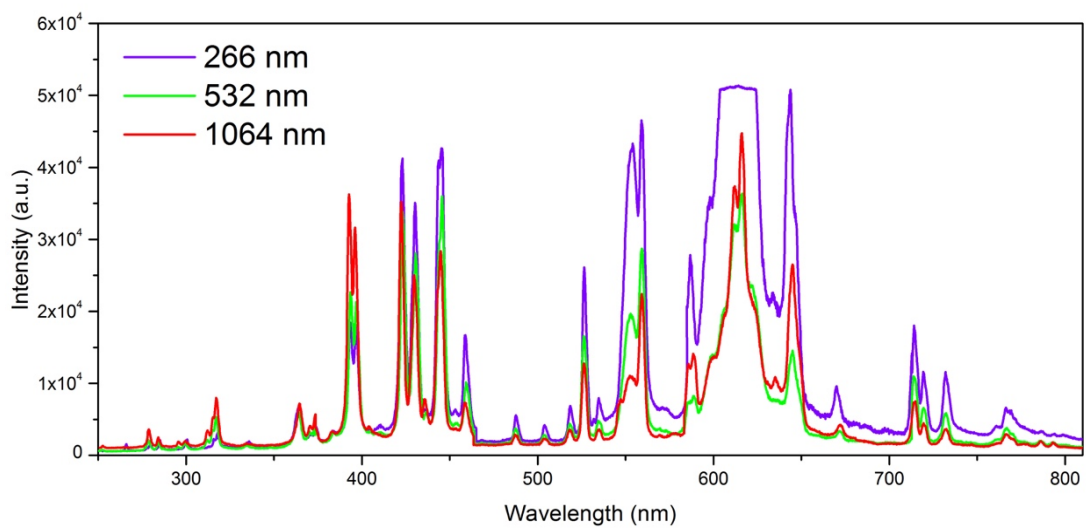

**Fig. 3. The total spectral emission on fresco sample induced by three laser wavelengths**

# 3. Results and discussions

## 3.1 Spectroscopic identification

When LIBS analytical technique is used, the samples composed by organic binders produce the molecular emission such as CN and $C_2$ which was more easily detected than inorganic carbon containers who need the chemical interaction to form such molecules. The spectra between 388.3 nm and 385.1 nm present the CN violet system ($B^2\Sigma^+$ - $X^2\Sigma^+$, $\Delta\nu = 0$) molecular emission. [27] From the Fig. 4., for these three laser wavelengths, the emission of CN band is clear for the binders of animal glue, oil and egg yolk. However, on the fresco technique and the casein sample, the CN band emission were not observed.

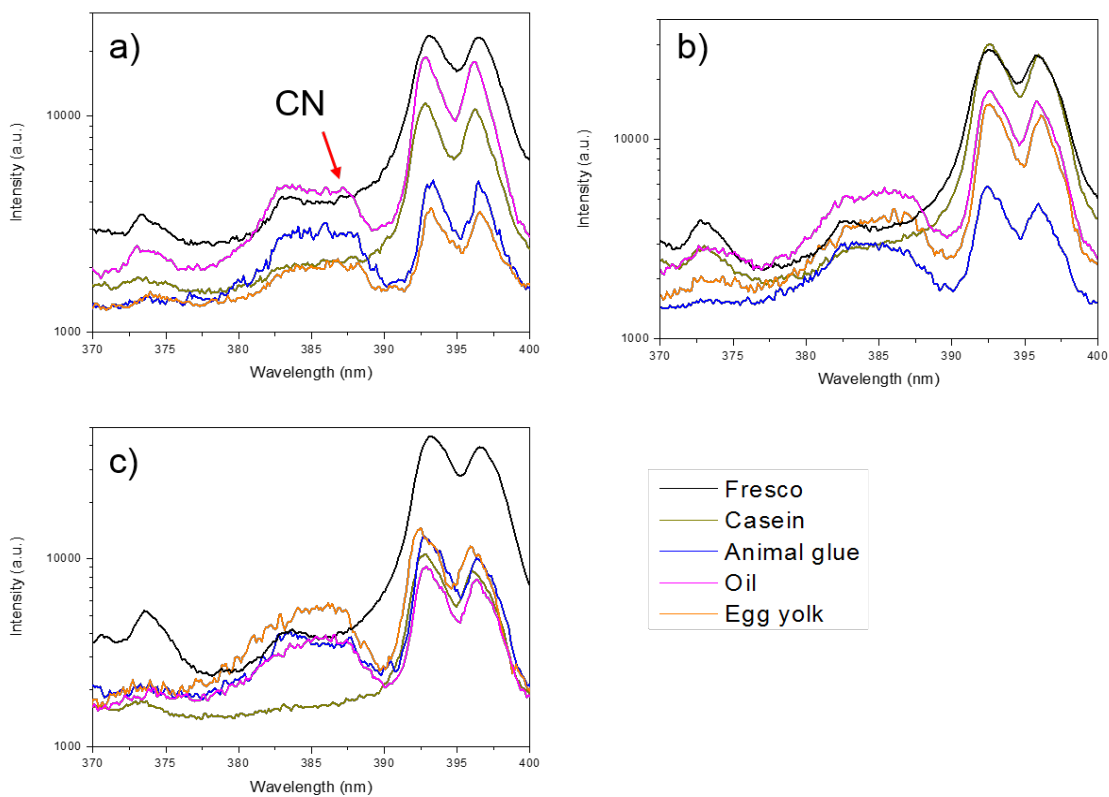

**Fig. 4.** The emission spectra including the CN violet system ($B^2\Sigma^+$ - $X^2\Sigma^+$) molecular emission spectrum for different painting techniques with a) 266 nm, b) 532 nm and c) 1064 nm laser excitation

The spectra between 505.6 nm and 516.5 nm present the $C_2$ Swan band system ($d^3\Pi_g$ - $a^3\Pi_u$, $\Delta v = 0$) emission. [27] From the Fig. 5., for these three laser wavelengths, the emission of $C_2$ band was more clear for the oil binder and an emission for egg yolk sample with 532 nm and 1064 nm. In the contrast, the $C_2$ band emission was not observed in other conditions.

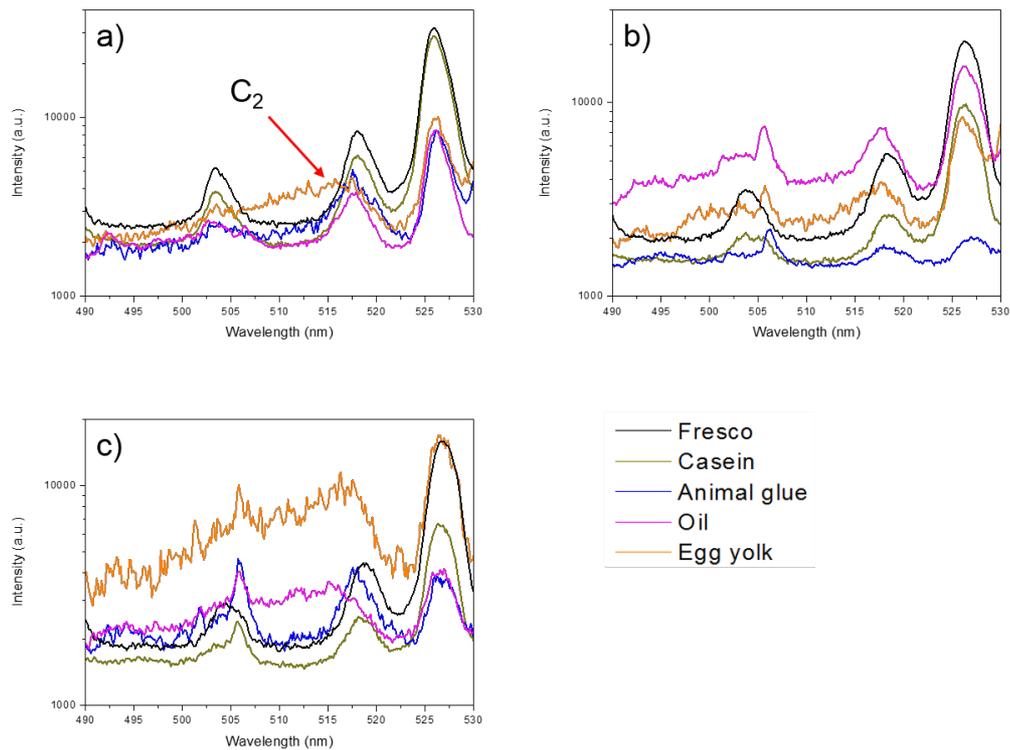

**Fig. 5. The emission spectra including the C$_2$ Swan system (d$^3\Pi_g$ - a$^3\Pi_u^+$) molecular emission spectrum for different painting techniques with a) 266 nm, b) 532 nm and c) 1064 nm laser excitation**

We resume the presence of the molecular emission induced by different laser wavelength on different samples in Table 4. The mark "-" represents that such emission was not detected in the spectra, "+" means that the corresponding emission was clearly observed but less intense than those with the mark "++". For 266 nm laser, the oil binder is easily distinguished by the strong emission of both CN and C$_2$ molecules. The animal glue and egg yolk can be separated from the others because of the different emission strength of CN molecules and without C$_2$ present. The fresco and casein sample could not be differentiated by the molecular emission. With the same method, for 532 nm and 1064 nm excitation, the animal glue, egg yolk and oil binder, they were also identified with combined comparison of the molecular emission of CN and C$_2$. Nevertheless, no molecular emission from fresco and casein binder, but according to their different natures, inorganic or organic, they can also be separated.

**Table 4**

The CN and $C_2$ emission for different samples. ("-": not emission, "++": Strong and "+": less intense than "++")

|  | 266 nm | | 532 nm | | 1064 nm | |
|---|---|---|---|---|---|---|
|  | CN | $C_2$ | CN | $C_2$ | CN | $C_2$ |
| Fresco | - | - | - | - | - | - |
| Animal glue | ++ | - | + | - | + | - |
| Casein | - | - | - | - | - | - |
| Egg yolk | + | - | + | + | ++ | +(+) |
| Oil | ++ | ++ | + | ++ | + | ++ |

## 3. 2 Chemometric identification

Although by comparing the molecular emission, the binders could be discriminated, but their identification must be difficult from only the intensity of emission spectra for a sample without any composition information. In order to more precisely discriminate the different painting techniques, we must treat the general information from the sample together. Therefore, the chemometric approach of principal component analysis is a useful tool to achieve this objective.

We first worked on the whole spectrum adding the first three wavelength range from 245 nm to 580 nm. The Fig. 6. shows the PCA results of this choosing wavelength range. For the three excitations, the fresco and casein binder samples were separated from each other, and obtained better results than compared to the spectral analysis. Whereas, the animal glue, oil and egg yolk could not be separated. As a results, to improve the sample discrimination, the investigation of the three spectral zone separately was used to compare and identify a suitable range.

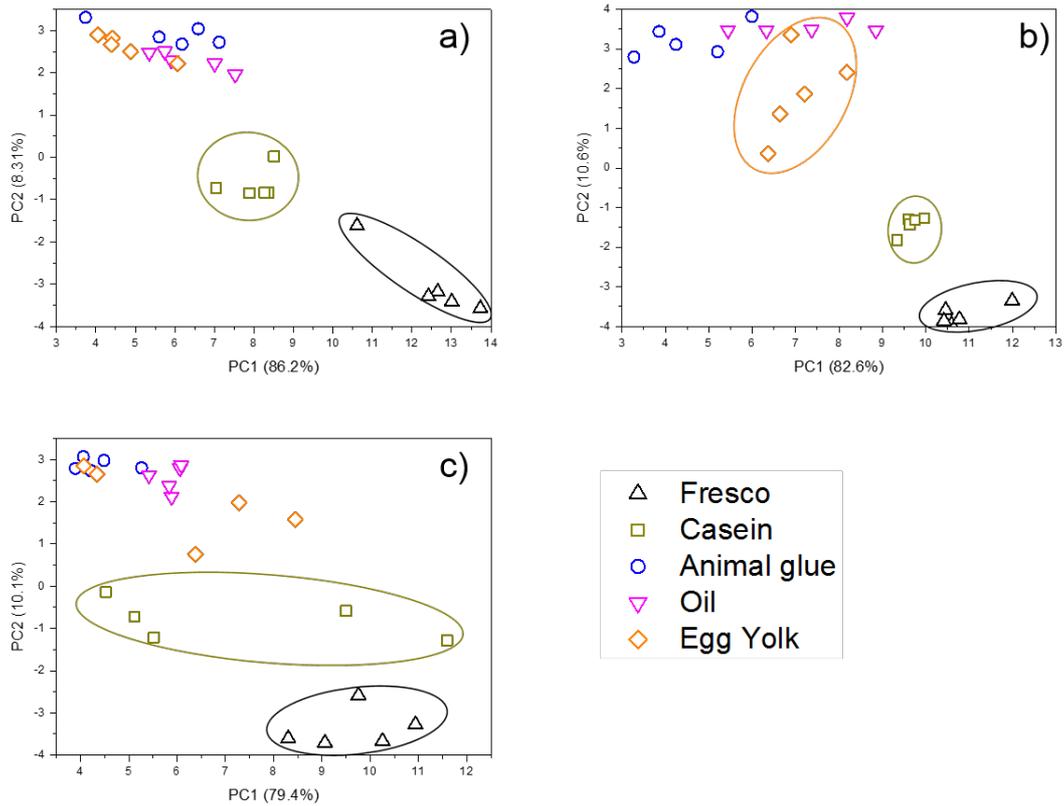

**Fig. 6.** Score plot of the two principal components, PC1 and PC2, of total variance between 245 and 580 nm with a) 266 nm, b) 532 nm and c) 1064 nm laser excitation

Studying different spectral zones separately, the first spectral zone is from 250 nm to 370 nm and the PCA results is shown in Fig. 7). In this spectral range, the emission is mainly from the elements, such as neutral carbon and neutral or ionized calcium. For 266-nm excitation (Fig. 7 a), the five techniques were well separated, for 532-nm excitation (Fig. 7 b), only the fresco and casein binder were identified and at last, for 1064-nm excitation (Fig. 7 c), the organic binder and inorganic fresco technique were discriminated. In this case, the 266 nm was more suitable for discriminating the binders.

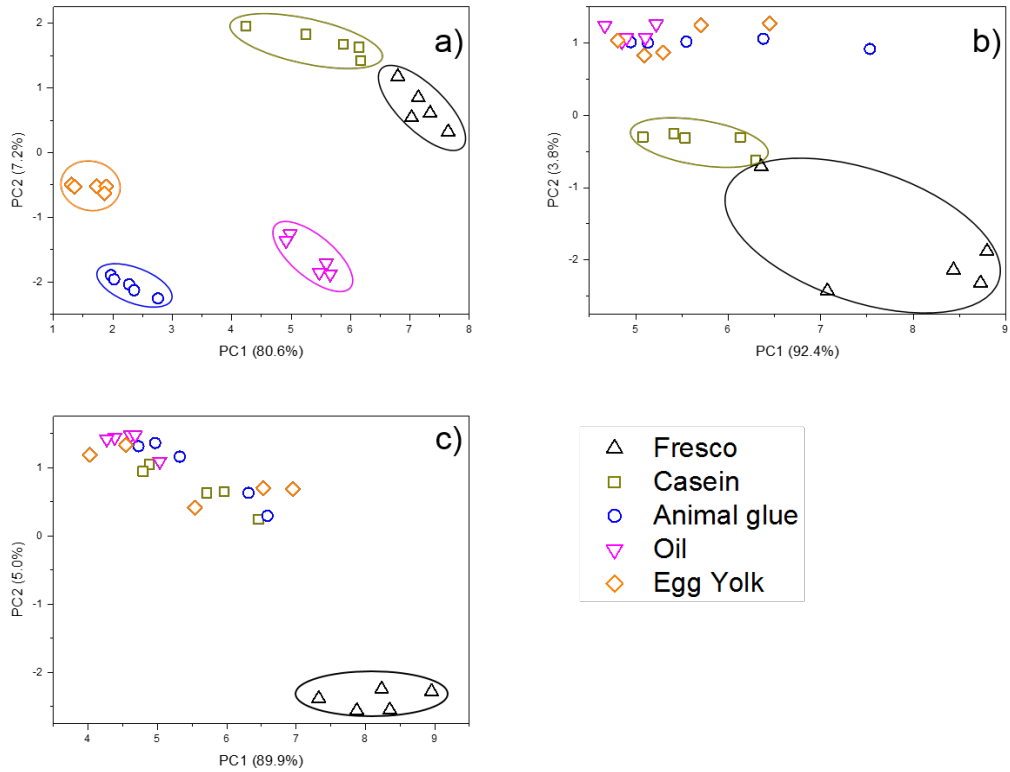

**Fig. 7.** Score plot of the two principal components, PC1 and PC2, of total variance between 250 and 370 nm with a) 266 nm, b) 532 nm and c) 1064 nm laser excitation

The second spectral zone studied is from 330 nm to 465 nm and the PCA results is shown in Fig. 8. In this spectral range, the emission was not only from the elements, such as neutral or ionized calcium and neutral mercury, but also including the CN band molecular emission. The PCA results demonstrated that the separation of different painting technique was achieved with all three exciting lasers.

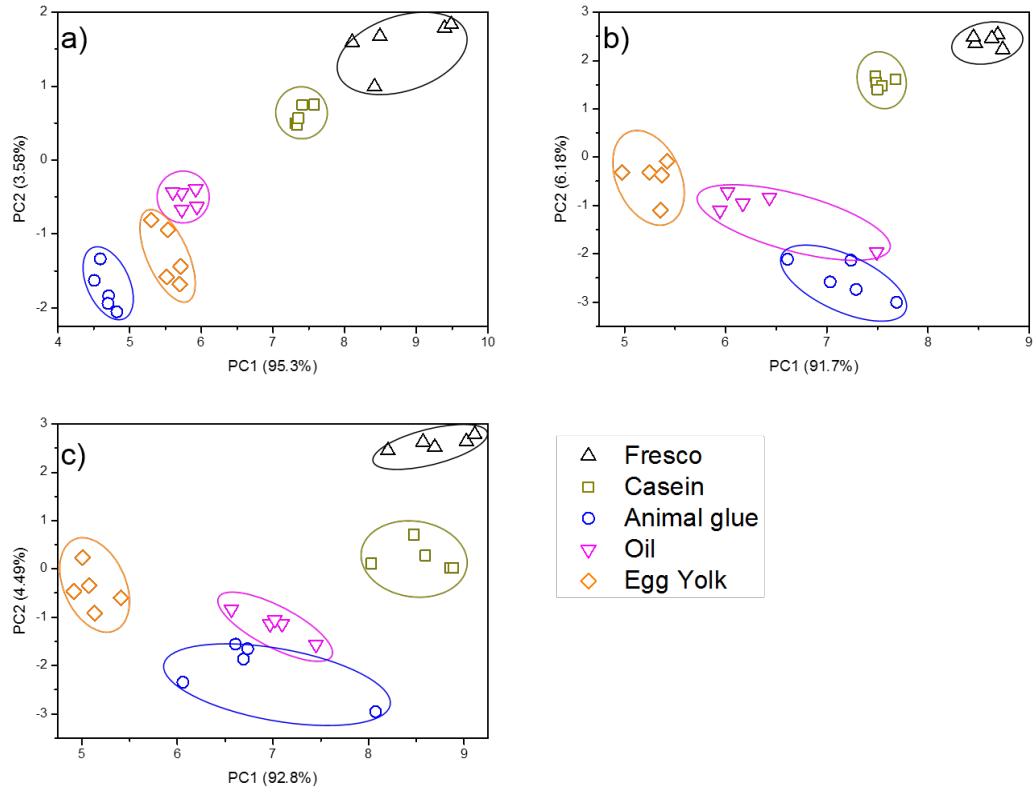

**Fig. 8.** Score plot of the two principal components, PC1 and PC2, of total variance between 330 and 465 nm with a) 266 nm, b) 532 nm and c) 1064 nm laser excitation

The last spectral zone studied is from 450 nm to 580 nm, and the PCA results of the first two principal components is shown in Fig. 9. In this spectral range, the emission was from the neutral calcium and/or the $C_2$ band molecular emission. Not like the spectral analysis, the binders were not separated by the presence of $C_2$ emission in PCA results. The fresco technique with more calcium is differentiated from the others. For 266-nm excitation (Fig. 9a), the oil binder was marked separately, which could be due to the contribution of $C_2$ emission.

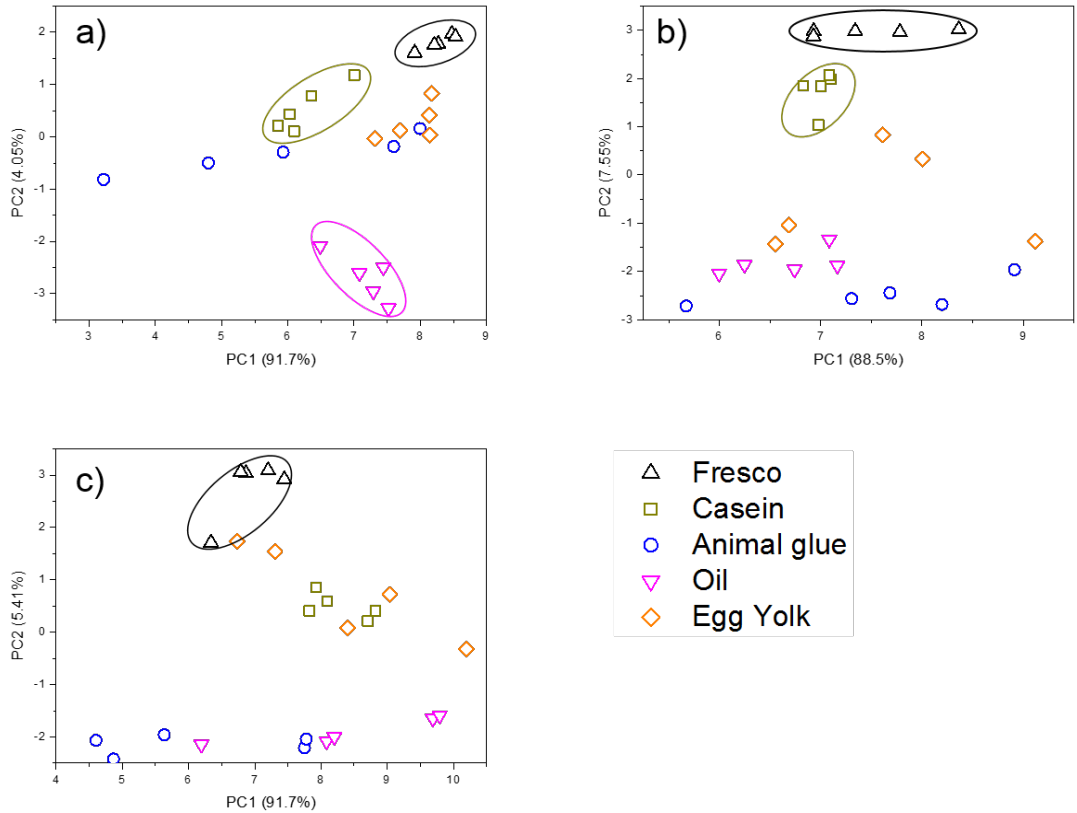

**Fig. 9.** Score plot of the two principal components, PC1 and PC2, of total variance between 450 and 580 nm with a) 266 nm, b) 532 nm and c) 1064 nm laser excitation

Further investigated was done with the third principal component shown in Fig. 10. The PCA results demonstrated that the separation of different painting technique could be achieved with all three exciting lasers from the second and third principal components. We can interpret that the emission spectra are similar in this zone and only a few differences among them, so that the first principal component is similar with which it is not possible to show the differences among the binders.

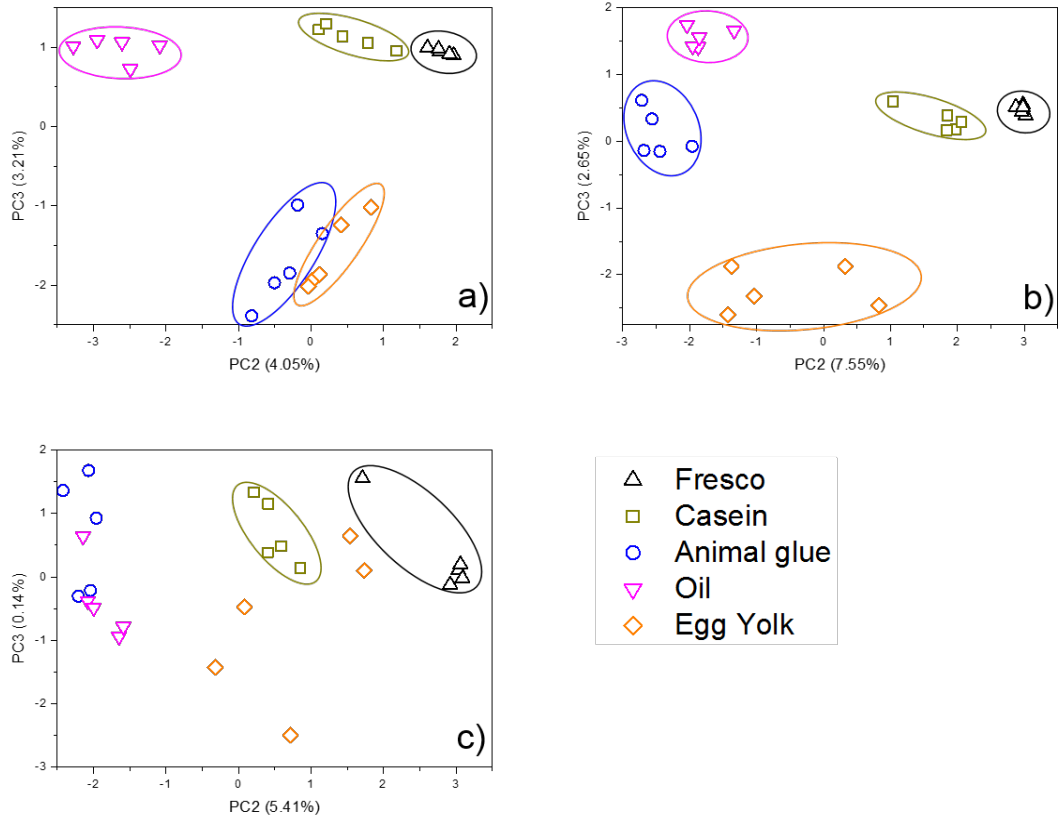

**Fig. 10.** Score plot of the two principal components, PC2 and PC3, of total variance between 450 and 580 nm with a) 266 nm, b) 532 nm and c) 1064 nm laser excitation

Summarizing the different ancient painting techniques (different binders) it could be discriminated by comparing the emission spectra or PCA. The 266 nm laser was more efficient to differentiate the binders than 532 nm or 1064 nm. As we showed, the influence of the laser energy can be neglected due to the adjustment of the laser pulse energy. Therefore, if the results become different with the organic binders, it must be due to the effects of different interaction between laser and the organic materials. The photon energies of lasers are 4.66, 2.33 and 1.17 eV for 266, 532 and 1064 nm, respectively. Comparing with the C-C bond energy is 3.59 eV [28], the chain of carbon is more easily broken by 266 nm and more CN and $C_2$ are produced. As a results, the stronger molecular emission has been observed. However, we observed also a stronger molecular emission with the 532 and 1064 nm. If the photons do not have enough energy, the molecular bands could also be broken into atoms or molecular fragments by electrons excited through heating effect in the plasma.

Therefore, the molecular emission can be generated at the beginning of the plasma life. In addition, the carbon in the plasma could also interact with the carbon or nitrogen atoms or ions to have CN and $C_2$ emission. The recombination in the plasma takes place under the certain temperature according to the dissociated energy of the bonds. Since the known dissociated energy for C-C bond and C-N bond are 6.27 eV and 7.72 eV [29], it must to study the electron temperature in the plasma in order to understand the different capacities shown by different laser wavelengths in separating the binders.

### 3.3 Estimation of the electron temperature in the plasma

### 3.3.1 Estimation of the "temperature" of the plasma induced on the fresco sample

According to the Boltzmann distribution law, the line ($\lambda_{ij}$) emission intensity $I_{ij}$ in the plasma is given by [30] Eq (1) with the influence of detection system:

$$I_{ij} = \frac{hc}{4\pi\lambda_{ij}} A_{ij} F_{ij} \frac{N}{U(T)} g_j \, exp\left(-\frac{E_j}{k_B T}\right)$$

(1)

where $h$ is the Planck constant, $c$ is the speed of light, $A_{ij}$ is the transition probability, $E_j$ and $g_j$ are the energy and degeneracy of the upper energy level $j$ respectively, $N$ is the number density of the atom and $k_B$ is the Boltzmann constant and $T$ is the plasma temperature, $U(T)$ is the partition function of the species at temperature $T$ and $F_{ij}$ is the factor concerning the optical collection system and the response of the ICCD camera. Then we have taken a second emission line with the transition for different upper level $E_n$ to lower level $E_m$ at the wavelength $\lambda_{mn}$ in order to remove the common parameters in the Eq (1) which we cannot get from the measurement.

By taking the ratio of the equations of these two emission lines and the natural logarithm, we get

$$\ln\left(\frac{I_{ij}\,\lambda_{ij}A_{mn}g_n}{I_{mn}\lambda_{mn}A_{ij}g_j}\right) = \ln\left(\frac{F_{ij}}{F_{mn}}\right) - \frac{E_j - E_n}{k_B T}$$

(2)

Therefore, we get the symbolic temperature

$$T^* = -\frac{1}{k_B T} + Const. = \frac{1}{E_j - E_n} \ln\left(\frac{I_{ij}\,\lambda_{ij}A_{mn}g_n}{I_{mn}\lambda_{mn}A_{ij}g_j}\right)$$

(3)

where

$$Const. = \frac{1}{E_j - E_n} \ln\left(\frac{F_{ij}}{F_{mn}}\right)$$

Therefore, the $T^*$ is an increasing function of the temperature so that the bigger value of $T^*$ represents a higher temperature of plasma. The lines used to estimate this "temperature" and their parameters are shown in the Table 5.

**Table 5**

The Ca lines and parameters are used in the estimation of temperature of the plasma [31], $\lambda_c$, $E_k$ and $A_{ki}g_{ki}$ are the center wavelength, the upper level energy and the product of transition probability and degeneracy of the transition, respectively.

| | | |
|---|---|---|
| $\lambda_c$ (nm) | 364.4 | 430.3 |
| $E_k$ (eV) | 5.3 | 4.78 |
| $A_{ki}g_{ki}$ (S$^{-1}$) | 2.49e8 | 6.8e8 |

The "temperature" results of the fresco sample are shown in Table 6 induced by different excitation wavelengths, it is coherent to the adjustment of laser pulse energy that the plasmas are in the same state because of the "same temperature".

**Table 6**

The "temperature" results for the fresco sample with different excitation wavelengths

| Excitation wavelength (nm) | 266 | 532 | 1064 |
|---|---|---|---|
| $T^*$ | -2.001 | -2.036 | -1.960 |

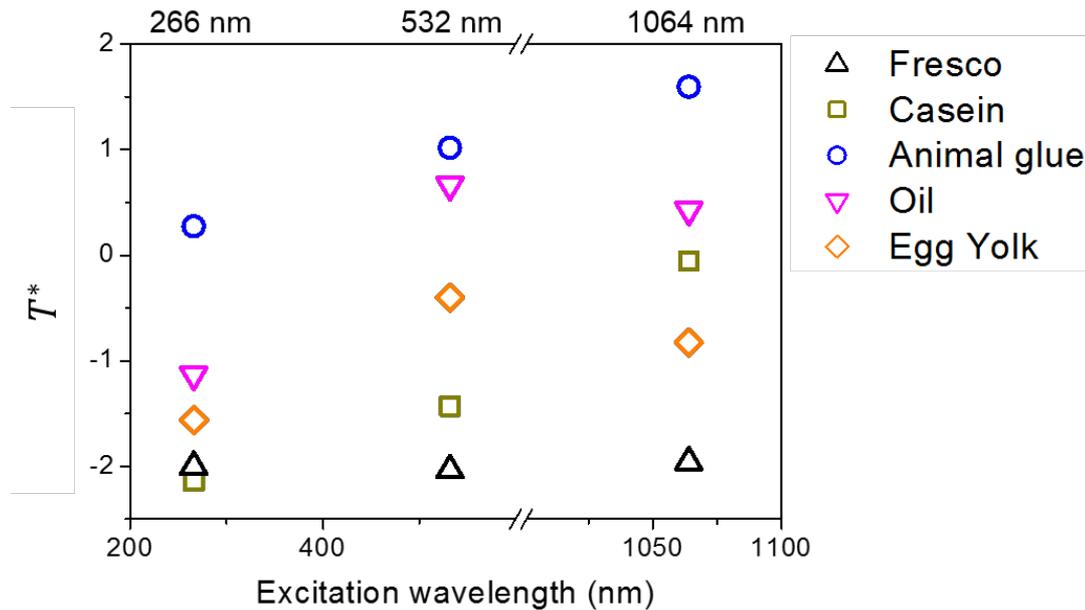

**Fig. 11.** The symbolic temperature $T^*$ excited by the three lasers with different wavelengths on the samples

### 3.3.2 Estimation of the "temperature" of the plasma induced on the sample with organic binders

For each different laser wavelengths, we applied the same pulse energy (in Table 2) on the samples of organic binders. The Fig. 11 shows the symbolic temperature $T^*$ excited by the three lasers on the organic binder samples. Also shown are the result of the fresco sample for comparison.

The casein and animal glue samples provided a higher "temperature" for the longer excitation wavelength. In the contrast, the oil and egg yolk samples had the highest temperature with the visible laser and the lowest one with UV laser. Referring to the chemical composition of these binders, the casein and the animal glue containing the acids protein and in the oil and egg yolk, the lipid presents, which provides the evident that the binder can be divided into two groups by the "temperature" as a function of the excitation wavelength of LIBS. For each group, the "temperature" were different: the animal glue and the oil samples had a relative higher temperature. By comparing the "temperature" of the plasma, the three laser wavelengths produced the different plasma corresponding to the different binders. For each sample, the lowest temperature with the 266 nm laser, we can interpret that with the high photon energy, the laser gave more energy to break the molecular bonds instead of matrix heat. At lower temperatures, it allowed more molecules forming. In addition, from the Fig. 3, the elemental emission show that it is strong enough to analyse the composition of the pictorial layers.

However, the long carbon chain is hard to direct dissociated for 532 nm and 1064 nm. As a result, with the photochemical mechanism of 532 nm laser, the hot electrons can be produced ant at the same time they can break the C-C bonds. Therefore, the temperature is higher with the sample of oil and egg yolk binders. The pure photothermal mechanism, because of strong heating effect with IR laser, for the casein and the animal glue samples was the highest temperature excited by 1064 nm.

## 4 Conclusion

In this work, we have studied the influence of ns-laser wavelength to discriminate painting techniques with single shot LIBS. And we have compared the emission of CN and $C_2$ band in the plasma induced by UV, VIS and IR laser pulses on different binder samples. With the condition of molecular emission presenting, the different binders could be discriminated by all three lasers and the 266 nm have a better performance. Using the chemometric approach of PCA to discriminate the different painting techniques the binder identification was possible using the results of PCA by LIBS, although the binders are organic and difficult to separate because of their similar elemental compositions. In deciding on the part of emission spectra including the most of information from the sample other than the whole spectral range. In the case of this work, the best range is tested is between 330 nm and 465 nm containing the neutral, ionic and molecular emission. The emission in this selected spectral range holds the most of different components from the sample and eliminated the repetitive and cumbrous emission, the calcium lines in the range from 450 nm to 580 nm for example. Furthermore, when it is difficult to decide the proper spectral range, the principal components with less variances than the first one are also able to provide a good discrimination instead. To understand the capacity of discrimination, the electron temperature in the plasma has been studied. The results showed that the photon energy played an important role in the formation of the plasma. With the plasma of the same electron temperature induced on the fresco sample, the temperatures in the plasma from the organic binders as a function of laser wavelength separate these samples into two groups: with or without long chain of carbon. The results in this paper can also be generalised to other inorganic pigment as a result of the similar reflectance in these laser wavelength range. Future work focus on the acquisition of the full information of composition of the pictorial layer, including the identification of both the pigments and binders or other restauration products. The work here allows us having a good solution of the choice of laser and of data treatment protocol for applying the LIBS technique in cultural heritages analysis.

## Acknowledgements

This work has been supported by the French Ministry Research Program EquipEx PATRIMEX and by IPERION CH project funded by the European Commission, H2020- INFRAIA-2014-2015, Grant No. 654028.